\documentclass[journal]{IEEEtran}

\pdfoutput=1
\usepackage{cite}

\ifCLASSINFOpdf
  \usepackage[pdftex]{graphicx}
  \DeclareGraphicsExtensions{.pdf,.jpeg,.png}
\fi
\usepackage{url}

\usepackage{hyperref}

\begin{document}
%
\title{Developing a Scenario-Based Video Game Generation Framework: Preliminary Results}
%
%
%

\author{Elif~Surer $^{(1)}$, Mustafa~Erkayao\u{g}lu $^{(2)}$, Zeynep Nur~\"{O}zt\"{u}rk $^{(3)}$, Furkan~Y\"{u}cel $^{(4)}$,  Emin~Alp~B{\i}y{\i}k $^{(5)}$, Burak~Altan $^{(6)}$, B\"{u}\c{s}ra~\c{S}enderin $^{(7)}$, Zeliha~O\u{g}uz $^{(8)}$,  Servet~G\"{u}rer $^{(9)}$, H.~\c{S}ebnem~D\"{u}zg\"{u}n $^{(10)}$\\
$^{(1)}$ Department of Modeling and Simulation, Graduate School of Informatics, Middle East Technical University, Ankara, Turkey\\
$^{(2)}$ Department of Mining Engineering, Middle East Technical University, Ankara, Turkey\\
$^{(3)}$ Department of Computer Engineering, Bilkent University, Ankara, Turkey\\
$^{(4)}$ Department of Modeling and Simulation, Graduate School of Informatics, Middle East Technical University, Ankara, Turkey\\
$^{(5)}$ Department of Modeling and Simulation, Graduate School of Informatics, Middle East Technical University, Ankara, Turkey\\
$^{(6)}$ Department of Modeling and Simulation, Graduate School of Informatics, Middle East Technical University, Ankara, Turkey\\
$^{(7)}$ Department of Modeling and Simulation, Graduate School of Informatics, Middle East Technical University, Ankara, Turkey\\
$^{(8)}$ Department of Psychology, Bilkent University, Ankara, Turkey\\
$^{(9)}$ Department of Mining Engineering, Middle East Technical University, Ankara, Turkey\\
$^{(10)}$ Colorado School of Mines, Brown Hall 268, CO 80401, USA\\
{\tt\small elifs@metu.edu.tr, emustafa@metu.edu.tr, nur.ozturk@ug.bilkent.edu.tr, furkanyucel.arch@gmail.com,
emin.biyik@metu.edu.tr, burak.altan@metu.edu.tr, busra.senderin@metu.edu.tr, zeliha.oguz@ug.bilkent.edu.tr, e119605@metu.edu.tr, duzgun@mines.edu}

}

\markboth{ENTERFACE'19, JULY 8TH - AUGUST 2ND, ANKARA, TURKEY}%
{ENTERFACE'19, JULY 8TH - AUGUST 2ND, ANKARA, TURKEY}
%



\maketitle

\begin{abstract}

Emergency training and planning provide structured curricula, rule-based action items, and interdisciplinary collaborative entities to imitate and teach real-life tasks. This rule-based structure enables the curricula to be transferred into other systematic learning platforms such as serious games ---games that have additional purposes rather than only entertainment. Serious games aim to educate, cure, and plan several real-life tasks and circumstances in an interactive, efficient, and user-friendly way. Although emergency training includes these highly structured and repetitive action responses, a general framework to map the training scenarios' actions, roles, and collaborative structures to game mechanics and game dialogues, is still not available. To address this issue, in this study, a scenario-based game generator, which maps domain-oriented tasks to game rules and game mechanics, was developed. Also, two serious games (i.e., Hospital game and BioGarden game) addressing the training mechanisms of Chemical, Biological, Radiological, Nuclear, and Explosives (CBRNe) domain, were developed by both the game developers and the scenario-based game generator for comparative analysis. The results show that although the game generator uses higher CPU time, memory usage, and rendering time, it highly outperforms the game development pipeline performance of the developers. Thus, this study is an initial attempt of a game generator which bridges the CBRNe practitioners and game developers to transform real-life training scenarios into video games efficiently and quickly.


\end{abstract}

\begin{IEEEkeywords}
Serious Games, Game Scenarios, Video Game Generator, CBRNe,  Emergency Training.\newline
\end{IEEEkeywords}

%
\IEEEpeerreviewmaketitle

\section{Introduction}
%
%
%
%
\IEEEPARstart{S}{erious} gaming \cite{IEEEhowto:santos}\cite{IEEEhowto:michael}, the umbrella term describing the video and board games having additional goals rather than only entertainment, is widely used in several domains such as health \cite{IEEEhowto:pirovano}, defense \cite{IEEEhowto:susi}\cite{IEEEhowto:crookall}, and education \cite{IEEEhowto:tinati}\cite{IEEEhowto:curtis}. CBRNe is an acronym for Chemical, Biological, Radiological, Nuclear, and Explosives, and recent research on this domain focuses on personnel training, emergency planning and organizing of field, tabletop, simulation, and serious gaming exercises for preparedness \cite{IEEEhowto:xvr}. 

One of the use cases of serious gaming in emergency planning is on firefighter training. In a study by Heldal \cite{IEEEhowto:heldal}, firefighter training was examined by using serious games and tools. To do so, qualitative questionnaires and observations on two use cases (i.e., ship evacuation in Baltic Sea and railway accident with cyanide leakage) were used to analyze the impacts of serious gaming on non-users. Results showed that serious games would be useful in emergency training situations, and in-depth training scenarios and evaluation methods were necessary.

In another study, Lukosch et al. \cite{IEEEhowto:lukosch} performed the steps of the traditional design process with the contributions of the end-users. The primary purpose of the study was to check if the simulations could be used to train situational awareness skills, and the end-users' participation simulation demonstrated the positive impact of using simulations. However, the main limitation of the study was not having a game-scenario based approach, and future research would focus on this aspect while creating virtual agents.

The use of virtual reality simulation was also a common topic in the literature. Ingrassia et al. \cite{IEEEhowto:ingrassia} focused on testing and comparing performances of 56 medical students during mass casualty triage in real-world and virtual reality (VR). The results showed that VR and live simulation were both useful in improving the accomplishments of the medical students. Ragazzoni et al. \cite{IEEEhowto:ragazzoni} also focused on VR training's medical aspect, where the objective was to increase the staff safety in life-or-death risks. Hybrid simulation for infection control and Ebola treatment were also successfully performed virtually, and the results demonstrated that awareness of the health personnel increased.

Serious gaming in CBRNe has been a recent topic, and there are some misinterpretations on the definitions and core concepts, such as the misuse of the words `game' or `simulation'. To overcome these misinterpretations, a pre-development survey was developed \cite{IEEEhowto:enoticesurvey} to be used before implementing the serious games of the European Network Of CBRN TraIning CEnters (eNOTICE) project \cite{IEEEhowto:enoticeweb}. In the pre-development survey, 24 questions were asked to the practitioners and experts of CBRNe under the following subgroups: 1) Participant's video gaming background, 2) Participant's knowledge on serious games, and 3) Participant's expectations on eNOTICE serious games. Results from 14 CBRNe professionals showed that the majority of the participants were highly positive on using serious games in CBRNe and provided open concepts, suggestions, and guidelines to develop serious games for CBRNe domain \cite{IEEEhowto:enoticenantes}.

In this study, a scenario-based game generator, which maps linear real-life scenarios to serious games with training objectives, was developed. The main focus of this study is the CBRNe domain, where the roles, tasks, and goals of the participants are clearly defined. The main objectives of the games are as follows: 1) Providing a tool for additional training, 2) Synergy building, and 3) Transporting a different domain and a new concept to the CBRNe community. Two of the games that were developed by both the scenario-based game generator and by the game developers were based on the real exercises that were performed during eNOTICE project's joint activities in Nimes (France) and Brussels (Belgium). The comparative analysis regarding CPU usage, rendering, memory usage, and game development pipeline on the generator-based and developer-based games was also performed.

\section{Materials and Methods}
In this section, details of the scenario-based game generator (Figures 1-3), two serious games that were developed by both the game designers and the game generator, and their evaluation are explained in detail. 

\subsubsection{Scenario and Task Definitions}The theme of the scenario, active players, location, and interaction mechanisms were collected in advance from the practitioners. Workflow and state diagrams were used to create a detailed scenario where different roles and entities communicate with each other. In the beginning, two different CBRNe scenarios that were based on real practices were mapped to the workflow and state diagram structures. One of the scenarios is based on the subset of the BioGarden exercise (i.e., linear version of the scenario in which the players do not change the flow of the events), which was held in June 2018 in Belgium as part of eNOTICE joint activities. The other scenario is based on the Nimes exercise, which was held in France in January 2018 again as part of eNOTICE joint activities.

Defined scenarios and interaction mechanisms were mapped to game ideas, linear game stories, and interaction mechanisms. Interaction mechanisms were converted into concrete tasks and user roles. A generic system, built on top of the initial scenario definitions, was conceptualized and implemented. Then, the generated system was fine-tuned with goals, feedback measurements, and score adaptations.

The scenarios of the exercises were designed by different institutions such as fire departments, research centers, and hospitals. Thus, breaking the scenarios down into actions and events was a crucial step so that the game mechanics, reward mechanisms, and scoring could be systematized. Also, different roles in the scenarios were assigned to different player types so that the active role of the player and the role of non-player characters (NPCs) were defined. 

\begin{figure}[!t]
\centering
\includegraphics[width=3.2in]{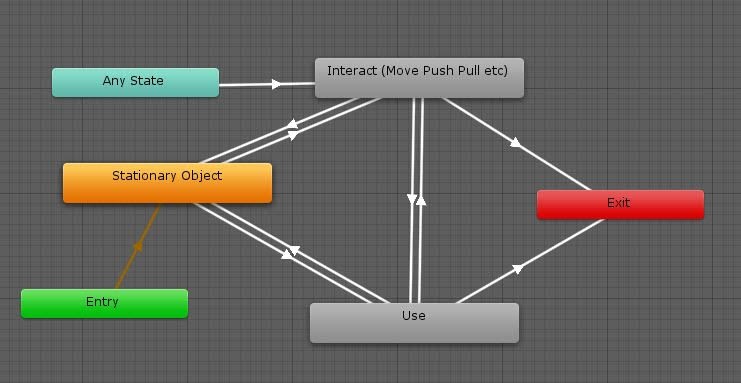}
\DeclareGraphicsExtensions.
\caption{Initial interaction mechanism including Entry and Exit States, a Stationary Object interacting (i.e., Move, Push, Pull, etc.) and using stationary objects.}
\label{scr8}
\end{figure}

\begin{figure}[!t]
\centering
\includegraphics[width=3.2in]{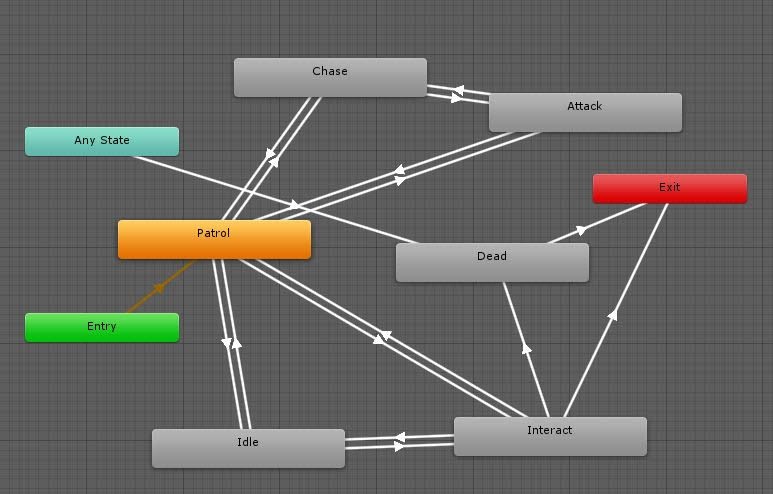}
\DeclareGraphicsExtensions.
\caption{In an attack scenario, chasing, attacking and interacting use cases are modeled.}
\label{scr10}
\end{figure}

\begin{figure}[!t]
\centering
\includegraphics[width=3.2in]{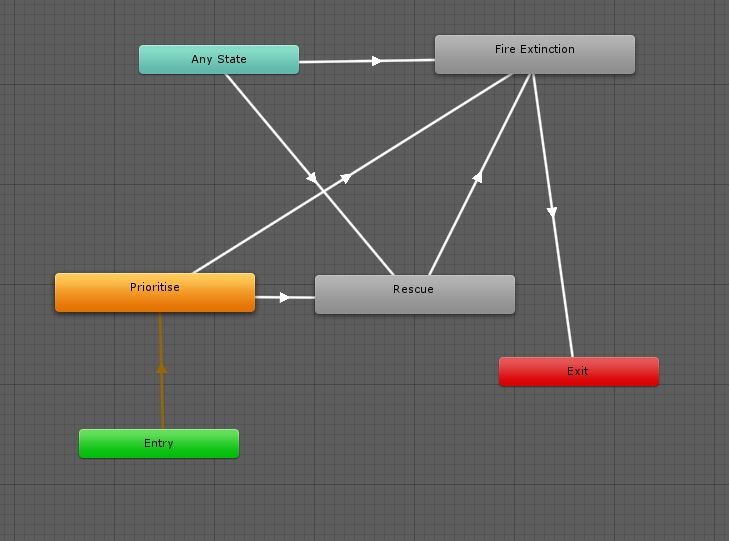}
\DeclareGraphicsExtensions.
\caption{Fire fighting scenario includes prioritizing the steps, extinguishing the fire and rescuing the affected people.}
\label{scr11}
\end{figure}

Before starting to implement the Hospital game, a detailed survey, which was briefly mentioned above, was conducted on 14 professionals in CBRNe field ---7 of them being game players \cite{IEEEhowto:enoticenantes}. The scope and the purposes of the study were as follows: 1) Learning the user's gamer profile, 2) Understanding the user's perspective on serious games, 3) Retrieving the expectations of the user, 4) Clarifying the differences between the video/serious games and simulations, and 5) Asking for suggestions. The initial results of the survey were positive and encouraging. The participants' gamer profiles involved playing strategy games and multiplayer games to learn new skills and to relieve stress. After the detailed analysis of the results, a tutorial mode was added to the initial game prototype.

\subsubsection{Scenario-Based Game Generator}
Scenario-based game generator is developed in combination with Unity software's Animator Controller tool and is composed of four different components: 1) Main Code, 2) Control Code, 3) Transition Code, and 4) User Interface. Main Code is where the state definitions and structures ---the definitions of the final consequences of the actions--- are initialized. Control Code works as a mechanism to form and map action methods and their related states. Transition Code is the link where the game generator works with Unity's Animator Controller. Finally, additional user interface mechanisms such as feedback, scores, and health points are added and grouped under the User Interface component (Figure 4). 

While developing the scenario-based game generator tool, the following steps are executed: 1) Creating an environment, 2) Defining state diagrams, 3) Creating animations, 4) Adding basic artificial intelligence (AI) to states, 5) Resolving player and NPC interactions and 6) Adding basic AI for interactions.

In this scenario-based game generator, only linear scenarios, where the decision making of players do not modify the outcomes of the actions, are implemented. The scenario-based game generator is used to generate the duplicates of Hospital and BioGarden games. First, the game scenarios are tested using simple 3D game objects such as cubes, spheres, and capsules (Figures 5 and 6). Then, after checking that the scenario works correctly, initial 3D game objects are automatically replaced with game assets using tag information of the assets.

\begin{figure}[!t]
\centering
\includegraphics[width=3.2in]{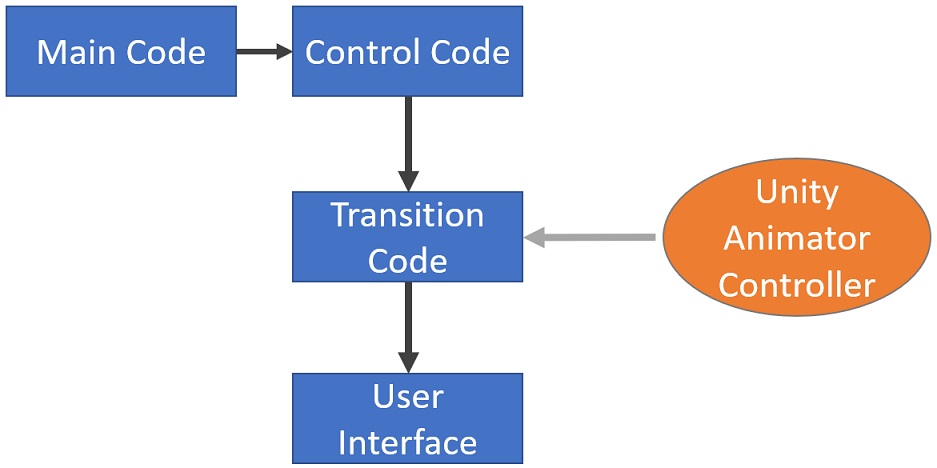}
\DeclareGraphicsExtensions.
\caption{Structure of the scenario-based game generator.}
\label{Flow}
\end{figure}

\subsubsection{Hospital Game}
Hospital Game is based on the Nimes scenario, which was performed during the eNOTICE joint activity in January 2018. The main purpose of this scenario is training medical staff for CBRNe circumstances. In this game, the player learns taking security measures such as using gloves, using masks, blocking the entrance of the hospital, and applying decontamination procedures. Players can play different roles, such as a doctor, nurse, and secretary. It is based on a linear scenario, and when the players make wrong choices, they lose game points (Figures 7 and 8).

This game was developed by a second-year Middle East Technical University (METU) Multimedia Informatics program student and the same game scenario was also given to the scenario-based game generator. The initial results of both environments were compared. In both versions of the games, Quadart's Hospital Lowply pack \cite{IEEEhowto:asseth}, which provides several realistic modular assets, was used.

\subsubsection{BioGarden Game}
BioGarden game is based on the eNOTICE joint activity, which was played in Belgium in June 2018. Although it had a nonlinear scenario, only linear parts of the scenario were implemented so that a comparison with the scenario-based game generator would be possible. In the scenario, there were different laboratories with different structures and responsibilities. The role-playing part was composed of decontamination, role assignment, and evaluation (Figures 9 and 10).

As in the case of the Hospital game, BioGarden game was developed by a second-year METU Multimedia Informatics program student, and the same game scenario was also given to the scenario-based game generator, and the initial results were compared. In both versions of the games, 3LB Games' Low Poly laboratory pack \cite{IEEEhowto:assetb}, which provides several realistic models, textures, and diffuse maps, was used.

\section{Results}

In this section, performance outcomes of the scenario-based game generator and the two serious games that were developed by the game developers were compared in terms of CPU usage, rendering time, memory usage, and game development pipeline. All the tests were performed on a laptop having Intel Core i7 9750HQ CPU, 16GB RAM, and NVIDIA GeForce GTX 1660TI graphics. 

Table I and Table II present the comparison on CPU usage (i.e. game generator's output vs. developer-based game), Table III and Table IV on memory usage, and finally, Table V and Table VI present a comparison using rendering parameters. The rendering profile used the SetPass Calls, Draw Calls, Total Batches, Triangles and Vertices as parameters. SetPass parameter is defined as ``the number of rendering passes'' \cite{IEEEhowto:renderer}, a Draw Call as a ``call to the graphics API to draw objects'' \cite{IEEEhowto:renderer} and Batch as a ``package with data that will be sent to the GPU'' \cite{IEEEhowto:renderer} on the Unity's Renderer Profiler page \cite{IEEEhowto:renderer}. 

\begin{figure}[!t]
\centering
\includegraphics[width=3.2in]{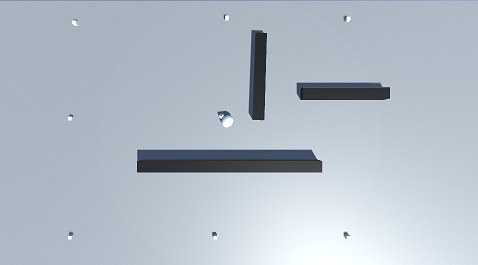}
\DeclareGraphicsExtensions.
\caption{Initial tests of the scenario-based game generator were performed on simple game objects.}
\label{scr1}
\end{figure}

\begin{figure}[!t]
\centering
\includegraphics[width=3.2in]{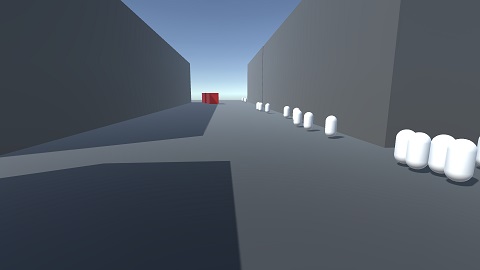}
\DeclareGraphicsExtensions.
\caption{Initial tests of the scenario-based game generator were performed on simple game objects such as cubes and capsules.}
\label{scr7}
\end{figure}

\begin{figure}[!t]
\centering
\includegraphics[width=3.2in]{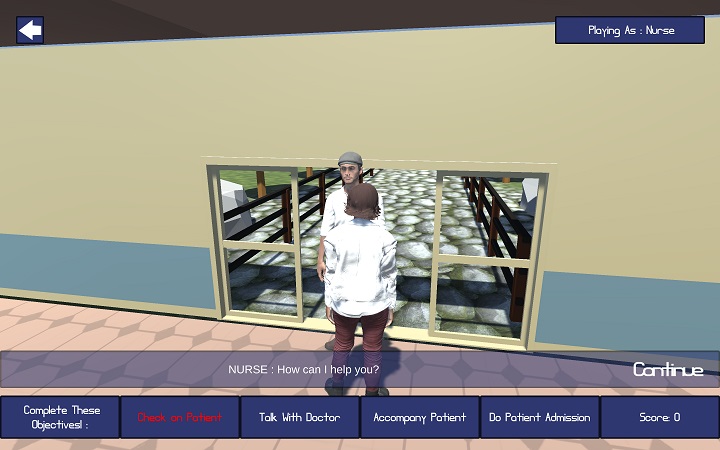}
\DeclareGraphicsExtensions.
\caption{Screenshots from the Hospital game and a demo of Dialogue menu.}
\label{scr13}
\end{figure}

\begin{figure}[!t]
\centering
\includegraphics[width=3.2in]{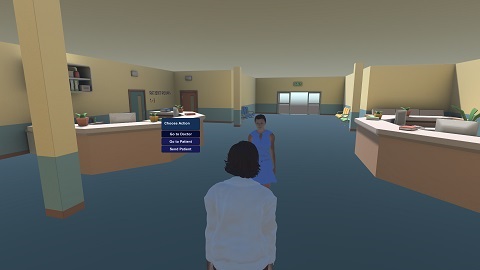}
\DeclareGraphicsExtensions.
\caption{Screenshots from the Hospital game and a demo of the interaction mechanism.}
\label{scr4}
\end{figure}

\begin{figure}[!t]
\centering
\includegraphics[width=3.2in]{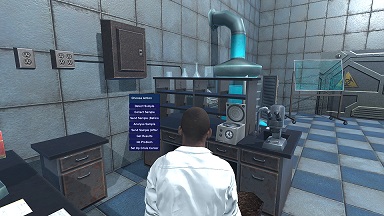}
\DeclareGraphicsExtensions.
\caption{The interior design of the Candestine lab from the BioGarden game and menu interactions.}
\label{scr2}
\end{figure}

\begin{figure}[!t]
\centering
\includegraphics[width=3.2in]{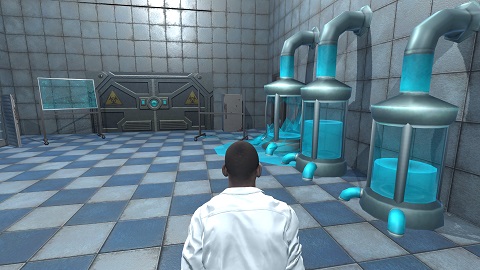}
\DeclareGraphicsExtensions.
\caption{The interior design of the Clandestine lab from the BioGarden game.}
\label{scr3}
\end{figure}

\begin{table}[!t]
\renewcommand{\arraystretch}{1.3}
\caption{CPU Usage of the Hospital Game Generated by the Scenario-Based Video Game Generator vs. Hospital Game}
\label{table_example}
\centering
\begin{tabular}{|c|c|c|}
\hline
CPU Usage & Hospital (Generator) &  Hospital Game \\
\hline
CPU & 9 ms & 4 ms\\
\hline
\end{tabular}
\end{table}

\begin{table}[!t]
\renewcommand{\arraystretch}{1.3}
\caption{CPU Usage of the BioGarden Game Generated by the Scenario-Based Video Game Generator vs. BioGarden Game}
\label{table_example}
\centering
\begin{tabular}{|c|c|c|}
\hline
Memory & BioGarden (Generator) &  BioGarden Game \\
\hline
CPU & 22.6 ms & 8.5 ms\\
\hline
\end{tabular}
\end{table}

\begin{table}[!t]
\renewcommand{\arraystretch}{1.3}
\caption{Memory Usage of the Hospital Game Generated by the Scenario-Based Video Game Generator vs. Hospital Game}
\label{table_example}
\centering
\begin{tabular}{|c|c|c|}
\hline
Memory & Hospital (Generator) &  Hospital Game \\
\hline
Used Total & 1.03 GB & 0.45 GB\\
\hline
Reserved Total & 1.32 GB & 0.63 GB\\
\hline
System Memory Usage & 1.96 GB & 1.38 GB \\
\hline
\end{tabular}
\end{table}

\begin{table}[!t]
\renewcommand{\arraystretch}{1.3}
\caption{Memory Usage of the BioGarden Game Generated by the Scenario-Based Video Game Generator vs. BioGarden Game}
\label{table_example}
\centering
\begin{tabular}{|c|c|c|}
\hline
Memory & BioGarden (Generator) &  BioGarden Game \\
\hline
Used Total & 0.62 GB & 0.28 GB\\
\hline
Reserved Total & 0.92 GB & 0.49 GB\\
\hline
System Memory Usage & 1.65 GB & 1.27 GB \\
\hline
\end{tabular}
\end{table}

\begin{table}[!t]
\renewcommand{\arraystretch}{1.3}
\caption{Rendering Results of the Hospital Game Generated by the Scenario-Based Video Game Generator vs. Hospital Game}
\label{table_example}
\centering
\begin{tabular}{|c|c|c|}
\hline
Rendering & Hospital (Generator) &  Hospital Game \\
\hline
SetPass Calls & 106 & 136\\
\hline
Draw Calls & 252 & 298\\
\hline
Total Batches & 217 & 243 \\
\hline
Triangles & 463.9K & 504.9K \\
\hline
Vertices & 319.5K & 361.1K \\
\hline
\end{tabular}
\end{table}

\begin{table}[!t]
\renewcommand{\arraystretch}{1.3}
\caption{Rendering Results of the BioGarden Game Generated by the Scenario-Based Video Game Generator vs. BioGarden Game}
\label{table_example}
\centering
\begin{tabular}{|c|c|c|}
\hline
Rendering &  BioGarden (Generator) &  BioGarden Game\\
\hline
SetPass Calls & 1826 & 2200\\
\hline
Draw Calls & 2584 & 3007\\
\hline
Total Batches & 2584 & 3007 \\
\hline
Triangles & 3.1M & 3.2M \\
\hline
Vertices & 2.3M & 2.4M \\
\hline
\end{tabular}
\end{table}

It took three weeks to develop and implement the scenario-based game generator. The most time-consuming part was the state transitions and handling the outcomes of the actions. After the game generator was built, it took three and a half hours to generate the Hospital game and four hours to generate the BioGarden game using the game generator.

The development of the original BioGarden game took 25 days in total: one week for the scenario clarification and role assignment; one week for the text-based decision mechanisms, nine days to finish the user interface and menus and two days to add the assets to the game.

The original Hospital game was first refactored from the initial prototype, which took one week. Then, it took another two weeks adapting the new scenario, merging different game modes, and dialogue generation.

\section{Discussion}

In this study, real-life exercise scenarios of two eNOTICE joint activities were developed by game developers as well as a scenario-based game generator ---developed during this study. All the frameworks used the same scenario-to-game mechanics mapping. All four versions of the games (i.e., developed by the game developer vs. generated by the game generator) were compared in terms of CPU usage, memory usage, rendering, and game development timeline perspectives. Scenario-based game generator's CPU usage, memory usage, and rendering time were higher when compared with the developer-based games. The reason for the game generator's higher resource usage was the complex structure of the scenario-based game generator, tag search, and communication with the Unity's Animator Controller. 

The rendering performance results of both versions of the games were very similar because the working principle of the game generator was not dependent on the visual contents of the games. Although the game generator used higher memory, CPU, and rendering time, its game development timeline efficiency highly outperformed the game developers (i.e., four hours vs. three weeks). This is a highly promising outcome that will enable further exercise scenarios to be mapped into games in a short period of time. This outcome can benefit the practitioners in two ways: 1) Visualizing the action-state diagrams of the exercise so that they can see the flaws or unassigned roles of their exercises, and 2) Having a rapid game prototype which becomes a fast, interactive testbed and training tool. 

The proposed game generator framework will be extended using state machines so that nonlinear scenarios can also be generated quickly. Also, use case scenarios will be adapted to VR environments \cite{IEEEhowto:surer}. The players will interact with their surroundings in the VR environment, achieve their goals, interact with other users and receive feedback regarding the success of their outcomes, which will enable us to build a detailed training environment where training scenarios can easily be modified and played in two different settings: on computers and using VR headsets. All the game versions (i.e., developer-based and game generator-based) will be played by the users, and a comparative analysis will be performed. Finally, the performance, technology acceptance, immersion,  and usability outcomes of the proposed system will be tested on participants and practitioners. Besides collecting game-related parameters such as interaction time and score, standard questionnaires on usability \cite{IEEEhowto:brooke} and technology acceptance model \cite{IEEEhowto:venkatesh} will also be applied to users. This tool will also be used to develop new prototype games for the future joint activities of the eNOTICE project till 2022.

\section{Conclusion}
In this study, a scenario-based video game generator, which targets the scenarios in CBRNe domain, was developed. This initial version of the game generator used linear scenarios that were based on the joint activities of the eNOTICE project. The effectiveness of the game generator was tested in comparison with two serious games, which were developed by the game developers. Even though the performance of the game generator lacked on the rendering, memory usage, and CPU usage aspects, it highly outperformed the game development pipeline of the game developers. This is a promising result that will enable the practitioners to visualize their scenarios while also generating prototype games rapidly so that the training of CBRNe personnel will be enriched. This current version of the game generator will be improved with usability tests, adaptation to VR and feedback of the CBRNe personnel so that an end-to-end and easy-to-use serious game generator for the CBRNe field will be provided.

\section*{Acknowledgments}

This study was developed during the 15$^{th}$ Summer Workshop on Multimodal
Interfaces (eNTERFACE'19) which was held at Bilkent University, Ankara,
Turkey between the dates of July 8 and August 2, 2019. This framework is
fully supported by European Network Of CBRN TraIning Centers (eNOTICE)
project funded under EU H2020 (Project ID: 740521). The authors would like
to thank Dr. Olga Vybornova (Center for Applied Molecular Technologies,
Université catholique de Louvain, Brussels, Belgium) and Prof. Gilles
Dusserre (IMT Mines Alès, Alès, Languedoc Roussillon, France) for their
help in scenarios of BioGarden and Hospital games, respectively. The
authors would also like to thank O\u{g}uzcan Erg\"{u}n (Multimedia Informatics,
Middle East Technical University, Ankara, Turkey) for his help and feedback on VR
instrumentation.

\ifCLASSOPTIONcaptionsoff
  \newpage
\fi



%

%

\begin{IEEEbiography}[{\includegraphics[width=1in,height=1.25in,clip,keepaspectratio]{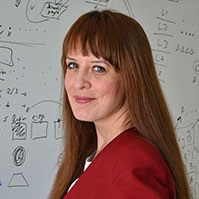}}]{Elif Surer}
Elif Surer received her Ph.D in Bioengineering in 2011 from the University of Bologna. She received her M.Sc. and B.Sc. degrees in Computer Engineering from Bo\u{g}azi\c{c}i University in 2007 and 2005, respectively. She is currently working as an Assistant Professor at the METU Graduate School of Informatics' Multimedia Informatics program. She is funded by the H2020 project eNOTICE as METU local coordinator as of September 2017 and also collaborates as a researcher in several interdisciplinary national and EU-funded projects. Her research interests are serious games, virtual/mixed reality, reinforcement learning and human and canine movement analysis.
\end{IEEEbiography}


\begin{IEEEbiography}[{\includegraphics[width=1in,height=1.25in,clip,keepaspectratio]{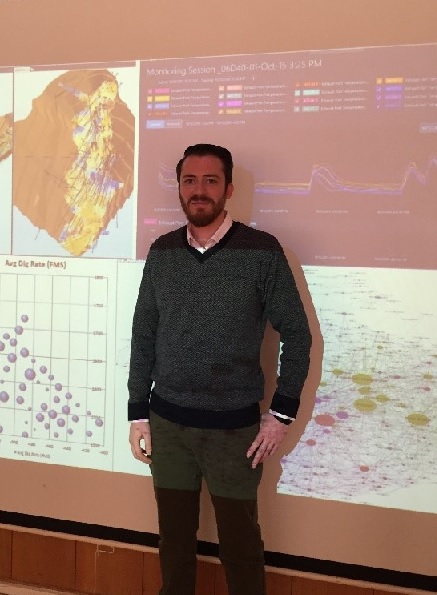}}]{Mustafa~Erkayao\u{g}lu}
Asst. Prof. Erkayao\u{g}lu, holding a PhD degree in mining engineering and a minor degree in Systems and Industrial Engineering earned from the University of Arizona, is an expert in business intelligence in the mining industry with a broad academic and practical experience. He has a proven track of teaching experience on mining engineering courses and his expertise also extends to project management, continuous improvement, and performance measurement skills acquired in the projects he worked as a Business Intelligence (BI) expert.
\end{IEEEbiography}

\begin{IEEEbiography}[{\includegraphics[width=1in,height=1.25in,clip,keepaspectratio]{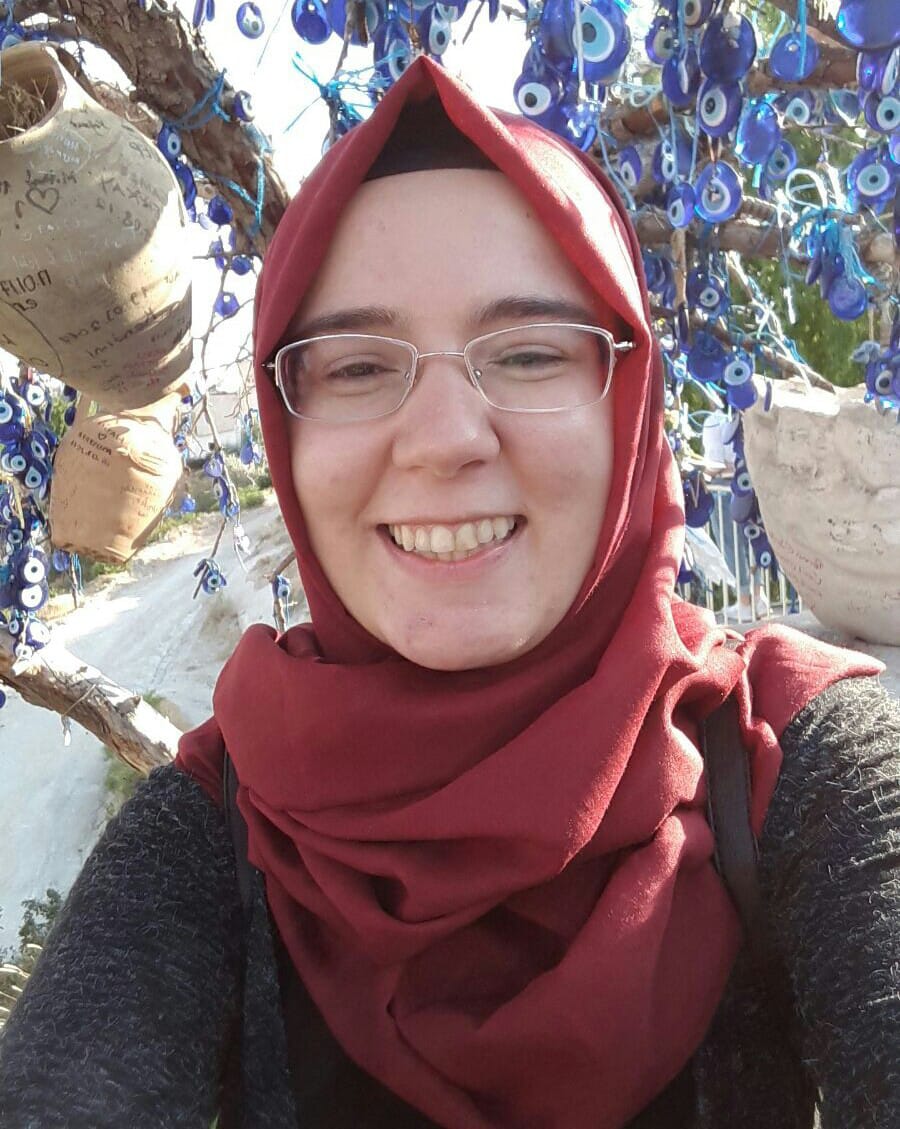}}]{Zeynep Nur~\"{O}zt\"{u}rk}
Zeynep Nur \"{O}zt\"{u}rk is studying on her Bachelor's degree in Computer Science at Bilkent University. During her Bachelor's degree, she studied basics of artificial intelligence and programming  games. Her focus is on game development with Unity, Unreal Engine and Java.
\end{IEEEbiography}

\begin{IEEEbiography}[{\includegraphics[width=1in,height=1.25in,clip,keepaspectratio]{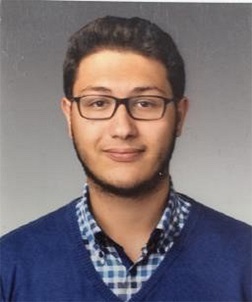}}]{Furkan~Y\"{u}cel}
Furkan Y\"{u}cel received his B.Arch. degree in Architecture from Bilkent University in 2019, where he studied generative algorithms and computational design in architecture. He is a first-year Master's student at METU Multimedia Informatics program.
\end{IEEEbiography}

\begin{IEEEbiography}[{\includegraphics[width=1in,height=1.25in,clip,keepaspectratio]{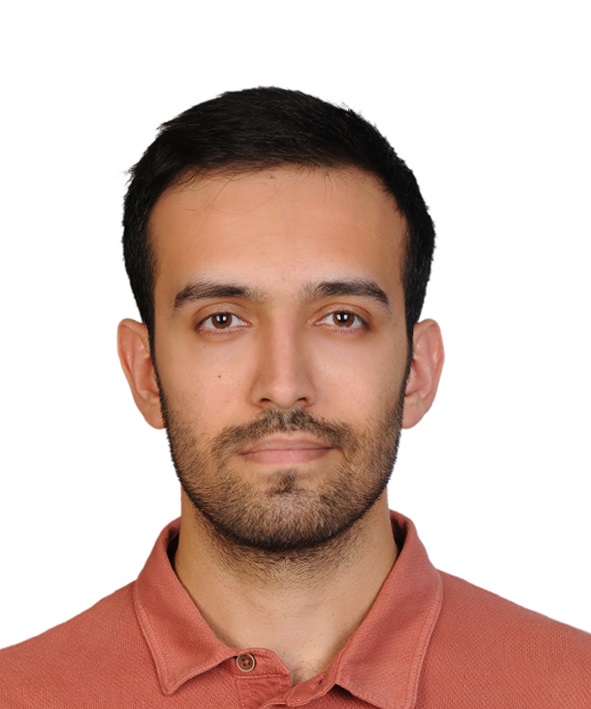}}]{Emin Alp~B{\i}y{\i}k}
Emin Alp B{\i}y{\i}k received his B.Arch degree in Architecture from Middle East Technical University (METU) in 2018, where he studied creative coding and generative design in architecture. He is a first-year Master's student METU Multimedia Informatics program. His research interests are game development, virtual/mixed reality, generative systems and multimedia arts.
\end{IEEEbiography}

\begin{IEEEbiography}[{\includegraphics[width=1in,height=1.25in,clip,keepaspectratio]{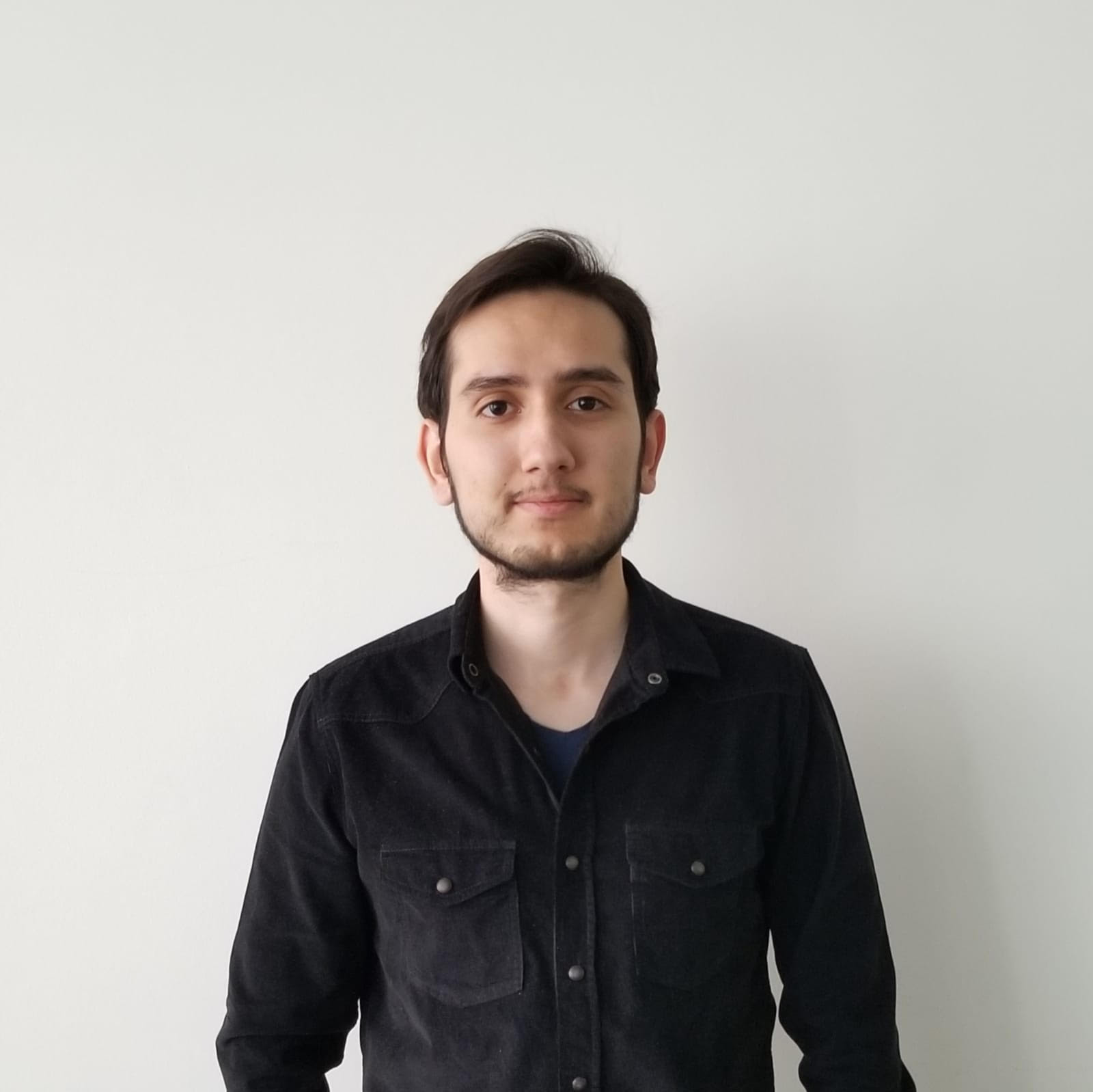}}]{Burak~Altan}
Burak Altan currently studies at METU Graduate School of Informatics' Multimedia Informatics program. He received his B.Sc. degree in Computer Engineering from Ba\c{s}kent University in 2017. He currently works at Ekin Teknoloji as a software engineer where his work includes front-end web development, Android development and database management. His research interests are serious games, virtual/mixed reality, and AAA game development.
\end{IEEEbiography}

\begin{IEEEbiography}[{\includegraphics[width=1in,height=1.25in,clip,keepaspectratio]{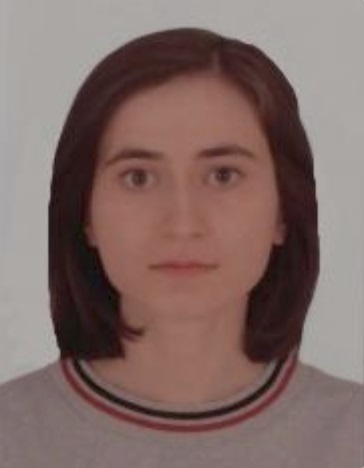}}]{B\"{u}\c{s}ra~\c{S}enderin}
B\"{u}\c{s}ra~\c{S}enderin is a second-year Master's student at METU Multimedia Infromatics program. Her current research focuses on serious games and virtual reality. She obtained her BS.c degree in Computer Engineering from Y{\i}ld{\i}r{\i}m Beyaz{\i}t University in 2018.
\end{IEEEbiography}

\begin{IEEEbiography}[{\includegraphics[width=1in,height=1.25in,clip,keepaspectratio]{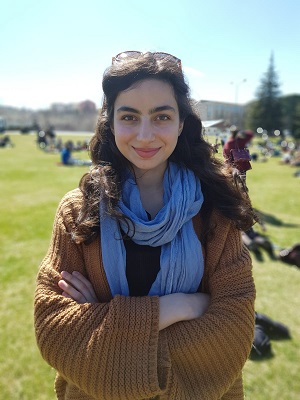}}]{Zeliha~O\u{g}uz}
Zeliha~O\u{g}uz is a second-year psychology student at Bilkent University. She has been working at Bilkent University's National Magnetic Resonance Research Center (UMRAM) as a research assistant. The research is about understanding gene-environment interactions by using neuroimaging, genetic and psychological analysis and multimedia arts.
\end{IEEEbiography}

\begin{IEEEbiography}[{\includegraphics[width=1in,height=1.25in,clip,keepaspectratio]{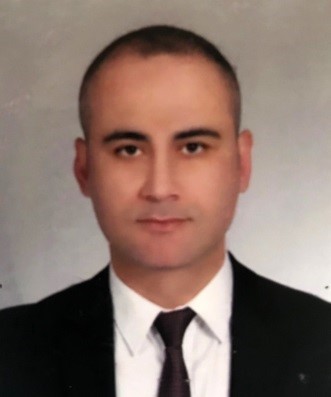}}]{Servet~G\"{u}rer}
Servet~G\"{u}rer received his BS.c degree in METU Mining Engineering Department and continues his Master's degree in the same department. After receiving his degree, he worked as a mining engineer in different companies and then became an occupational safety specialist. Since 2011, he has been working as a labor
inspector at the Ministry of Family, Labor and Social Services. He is a member of TMMOB Chamber of Mining Engineers and Labor Inspectors Association.
He served as the 44$^{th}$ board member of the Chamber of Mining Engineers. He works in the fields of 3D modeling, animation, game development, virtual reality. 
\end{IEEEbiography}

\begin{IEEEbiography}[{\includegraphics[width=1in,height=1.25in,clip,keepaspectratio]{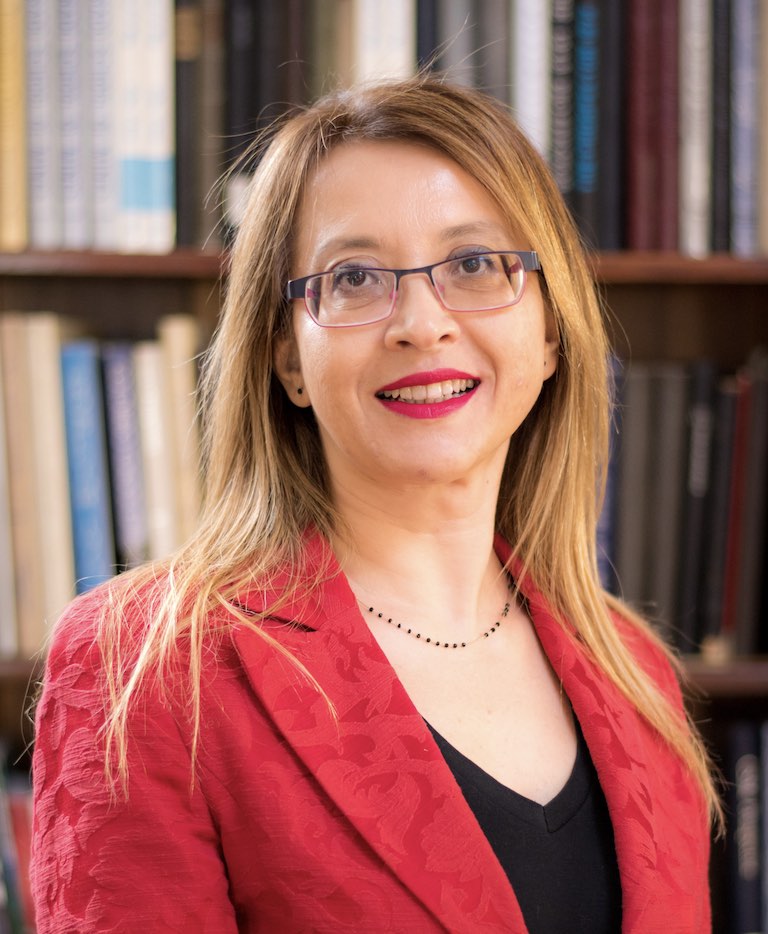}}]{H. \c{S}ebnem~D\"{u}zg\"{u}n}
Dr. H. \c{S}ebnem D\"{u}zg\"{u}n is Professor and Fred Banfield Distinguished Endowed Chair in Mining Engineering at Colorado School of Mines, Golden, USA. She also holds a joint appointment in the Department of Computer Science at Mines. Her recent research areas involve quantitative risk and resilience assessment for mining hazards and geohazards, big data analytics, Earth observation in geosciences, virtual/augmented/mixed reality (VR/AR/MR) and serious gaming for technical training and collaborative decision making.
\end{IEEEbiography}

\end{document}